\documentclass[aps,pra,graphicx,groupedaddress,reprint,showpacs]{revtex4-1}
\usepackage{amsmath}
\usepackage{amssymb}
\usepackage{graphicx}
\usepackage{hyperref}
\usepackage{tensor}
\usepackage{txfontsb} 
\usepackage{multirow}
\DeclareMathAlphabet{\mathcal}{OMS}{cmsy}{m}{n} 
\hypersetup{
    colorlinks=true,
    linkcolor=blue,
    citecolor=blue,
    filecolor=blue,
    urlcolor=blue,
}
\usepackage[braket]{qcircuit}

\begin{document}

\title{Transversal Clifford gates on folded surface codes}

\author{Jonathan E. Moussa}
\email{godotalgorithm@gmail.com}
\affiliation{Center for Computing Research, Sandia National Laboratories, Albuquerque, New Mexico 87185, USA}

\begin{abstract}
\bigskip
 \centerline{\begin{minipage}{0.79\textwidth}
\ \ \ \ Surface and color codes are two forms of topological quantum error correction in two spatial dimensions with complementary properties.
Surface codes have lower-depth error detection circuits and well-developed decoders to interpret and correct errors,
 while color codes have transversal Clifford gates and better code efficiency in the number of physical qubits needed to achieve a given code distance.
A formal equivalence exists between color codes and folded surface codes, but it does not guarantee the transferability of any of these favorable properties.
However, the equivalence does imply the existence of constant-depth circuit implementations of logical Clifford gates on folded surface codes.
We achieve and improve this result by constructing two families of folded surface codes with transversal Clifford gates.
This construction is presented generally for qudits of any dimension.
The specific application of these codes to universal quantum computation based on qubit fusion is also discussed.
\bigskip
\end{minipage}}
\end{abstract}

\maketitle

\section{Introduction}

The science of quantum error correction is maturing,
 with much of its development now shifting from general theory and properties of quantum codes
 to focused studies of the logical operation, decoding, and performance of specific codes and code families of particular importance.
With severe geometric restrictions expected of the layouts and interactions of qubits,
 special attention is directed at topological codes in two spatial dimensions
 that require only nearest-neighbor gate operations between qubits on a regular lattice.
The simplest families of these codes are surface codes \cite{surface_code} and color codes \cite{color_code}.
Surface codes are especially well understood, with resource estimates for the operation of quantum computers based on them \cite{quantum_resources}.

Surface codes are popular because of their simple structure and favorable attributes.
Data and ancilla qubits are arranged in a checkerboard pattern on a square lattice.
A surface code is maintained by repeatedly measuring the Pauli operators that generate its stabilizer group to detect errors.
All ancilla qubits are simultaneously initialized,
 then entangled with their four neighboring data qubits, and finally measured.
The low depth of this error detection circuit combined with efficient decoding of errors
 based on minimum weight perfect matching results in high error thresholds for faulty gate operations \cite{surface_threshold}.
Logical operations on a surface code are decomposed into Clifford and $T$ gates, 
 which are more difficult to implement than the error detection cycle.
Established implementations \cite{surface_operations} require code deformation for Hadamard ($H$) and \textsc{cnot} gates
 and resource states for phase ($S$) gates.
Magic state distillation \cite{magic_state_distillation} is used to implement $T$ gates, which has a very high resource overhead.
Much research is now devoted to reducing these overheads.

Color codes were developed to simplify logical operations with strongly transversal single-qubit Clifford gates \cite{color_code}.
These logical gates are implemented by applying the same physical gate to every data qubit in the code.
Another benefit is that fewer data qubits are needed to implement a logical qubit with distance $D$
 in a planar color code ($\tfrac{1}{2} D^2 + D - \tfrac{1}{2}$) relative to a planar surface code ($D^2$) \cite{color_efficiency}.
However, some stabilizer group generators of color codes have higher weight than in surface codes,
 which causes deeper error detection circuits and ancilla qubits that must interact with more than four neighboring data qubits.
The higher difficulty of detecting and decoding errors
 causes lower error thresholds for color codes compared with surface codes \cite{color_threshold}.
Experimental comparison of surface \cite{surface_experiment} and color \cite{color_experiment} codes is at an early stage.
With similar error rates, surface codes will reach their threshold before color codes.

A formal equivalence between surface and color codes has been established recently \cite{fold_surface}.
A color code on a triangle can be transformed into a surface code on a square by combining 
 local unitary operations with addition and removal of qubits
 and a final global unfolding operation.
The practical value of such a transformation is unclear.
Changes to the weight and locality of stabilizer generators and the number of data qubits
 will alter code performance, which restricts the mapping of useful properties between surface and color codes.
However, the transversal single-qubit Clifford gates of color codes must inevitably map to local unitaries on folded surface codes.
The details of this mapping have not yet been elucidated.

In this paper, we construct folded versions of two standard families of surface codes that are square or diamond segments of the toric code \cite{surface_families}.
Surprisingly, both code families admit transversal implementations of single-qubit Clifford gates,
 but at the cost of being reduced to the same low code efficiency of $2 D^2 - 2 D + 1$ data qubits per logical qubit for distance $D$.
We present this result for qudits rather than just qubits to extend
 the recent construction of transversal Clifford gates for qudit color codes in two spatial dimensions \cite{qudit_color}.
To simplify this presentation, we consider the specific case of $D = 5$ in detail
 and explain the straightforward extension to other $D$ value.

The paper is organized as follows.
A preliminary overview of quantum error correction with an emphasis on qudits and
 code symmetries is presented in Sec. \ref{prelim}.
The construction of folded surface codes and the qualitative effects of folding on the detection and decoding of errors are discussed in Sec. \ref{fold}.
The application of the folded cone code to universal quantum computation based on qubit fusion is considered in Sec. \ref{fusion}.
We conclude with the open problems that must be solved for these ideas to attain their full potential in Sec. \ref{conclude}.

\section{Preliminaries\label{prelim}}

The concepts and notation used in this paper closely follow the standards of quantum information theory \cite{textbook} with a few notable exceptions.
Several results are presented generally for qudits in $d$ dimensions, rather than specifically for the $d=2$ qubit case.
The purpose of this general presentation is to build a topological quantum error correction foundation compatible
 with the recently proposed concept of qubit fusion \cite{qubit_fusion},
 which combines Clifford gates on qubits and $d=4$ qudits to form a universal gate set.
To construct surface codes with transversal Clifford gates,
 we identify enabling code symmetries that are compatible with modified versions of the surface codes.

\subsection{Qudit notation\label{qudits}}

We define qudit notation with some purposeful ambiguity.
Qudit Pauli and Clifford operators are defined for an arbitrary dimension $d$,
 but we omit a specification of $d$ on their labels.
Unless specified further, all qudit equations are valid for all $d$.
In all quantum circuits, qudit wires are labeled with $d$.
Qubit wires are unlabeled, and $d=4$ qudit wires are labeled with a slash to signify their equivalence to a pair of qubit wires.

While not unique, there is a conventional generalization of qubits to qudits.
Every qudit has a set of computational basis states $|x\rangle$ for integers $x$ from $0$ to $d-1$.
We assume the ability to prepare physical qudits in $|0\rangle$ and perform measurements in their computational basis.
The qudit Pauli group is generated by $\{\omega^{1/2}, X , Z \}$, with generators defined as
\begin{align} 
 \omega &= \exp(2 \pi i / d), \label{qudit_phase} \\
 X &= \sum_{x=0}^{d-1} | (x+1) \bmod d \rangle \langle x |, \label{qudit_X} \\
 Z &= \sum_{x=0}^{d-1} \omega^x |x \rangle \langle x| . \label{qudit_Z}
\end{align}
While not a useful concept for $d>2$, the qudit generalization of the Pauli $Y$ is $Y = -\omega^{-1/2} X Z$ rather than $Y = \omega^{1/2} X Z$.
The single-qudit Clifford group is generated up to a phase by \cite{qudit_clifford}
\begin{align}
 H &= \sum_{x=0}^{d-1} \sum_{y=0}^{d-1} \frac{\omega^{xy}}{\sqrt{d}} | x \rangle \langle y | , \\
 S &= \sum_{x=0}^{d-1} \omega^{(x-d-2)x/2} |x \rangle \langle x| , \label{S_gate}
\end{align}
 and $Z$.
For $d = 2$ only, $Z$ and $S$ are dependent, with $Z = S^2$.
The multi-qudit Clifford group is generated up to a phase by including a controlled-$X$ gate between all pairs of qudits \cite{qudit_clifford},
\begin{equation}
 CX = \sum_{x=0}^{d-1} \sum_{y=0}^{d-1} | x \rangle \langle x | \otimes | (x+y) \bmod d  \rangle \langle y | ,
\end{equation}
 where an $X^n$ operation on the second qudit is controlled by the value $n$ of the first qudit.

The conjugation of qudit Pauli operators by qudit Clifford operators
 has a description that is independent of $d$ and can be decomposed
 into elementary conjugations of Pauli generators by Clifford generators.
The essential conjugation rules are
\begin{align}
 Z X Z^\dag &= \omega X \notag \\
 H X H^\dag &= Z \notag \\
 H Z H^\dag &= X^\dag \notag \\
 S X S^\dag &= -\omega^{-1/2} X Z \notag \\
 S Z S^\dag &= Z \notag \\
 CX  ( X \otimes I ) CX^\dag &= X \otimes X \notag \\
 CX ( I \otimes X ) CX^\dag &= I \otimes X \notag \\
 CX ( Z \otimes I ) CX^\dag &= Z \otimes I \notag \\
 CX ( I \otimes Z ) CX^\dag &= Z^\dag \otimes Z . \label{conjugation}
\end{align}
To achieve this $d$ independence, we use a nonstandard choice of $S$ in Eq. (\ref{S_gate}).
It is related to the standard choice \cite{qudit_clifford} by
\begin{equation}
 S_{\mathrm{standard}} = \left\{  \begin{array}{ll} Z^{1+d/2} S , & d \ \mathrm{even}, \\ Z^{(1+d)/2} S , & d \ \mathrm{odd}. \end{array} \right.
\end{equation}
The difference between Eq. (\ref{conjugation}) and qubit-specific rules is the distinction between operators and their Hermitian conjugates.
All of these operators are Hermitian in the qubit case, making this distinction unnecessary.

While qudits are not as frequently studied as qubits, many standard results for qubits have a straightforward extension to qudits.
For example, qudit stabilizer circuits are composed of qudit state preparation and measurement in the computational basis and qudit Clifford gates.
Just as in the qubit case, qudit stabilizer circuits can be efficiently classically simulated \cite{qudit_stabilizer}.
Stabilizer codes have also been extended to qudits \cite{qudit_QEC}.
When combined, these two results facilitate the computational study of quantum error correction on qudits.

\subsection{Qubit fusion\label{fusion_subsection}}

\begin{figure}[!t]
\includegraphics{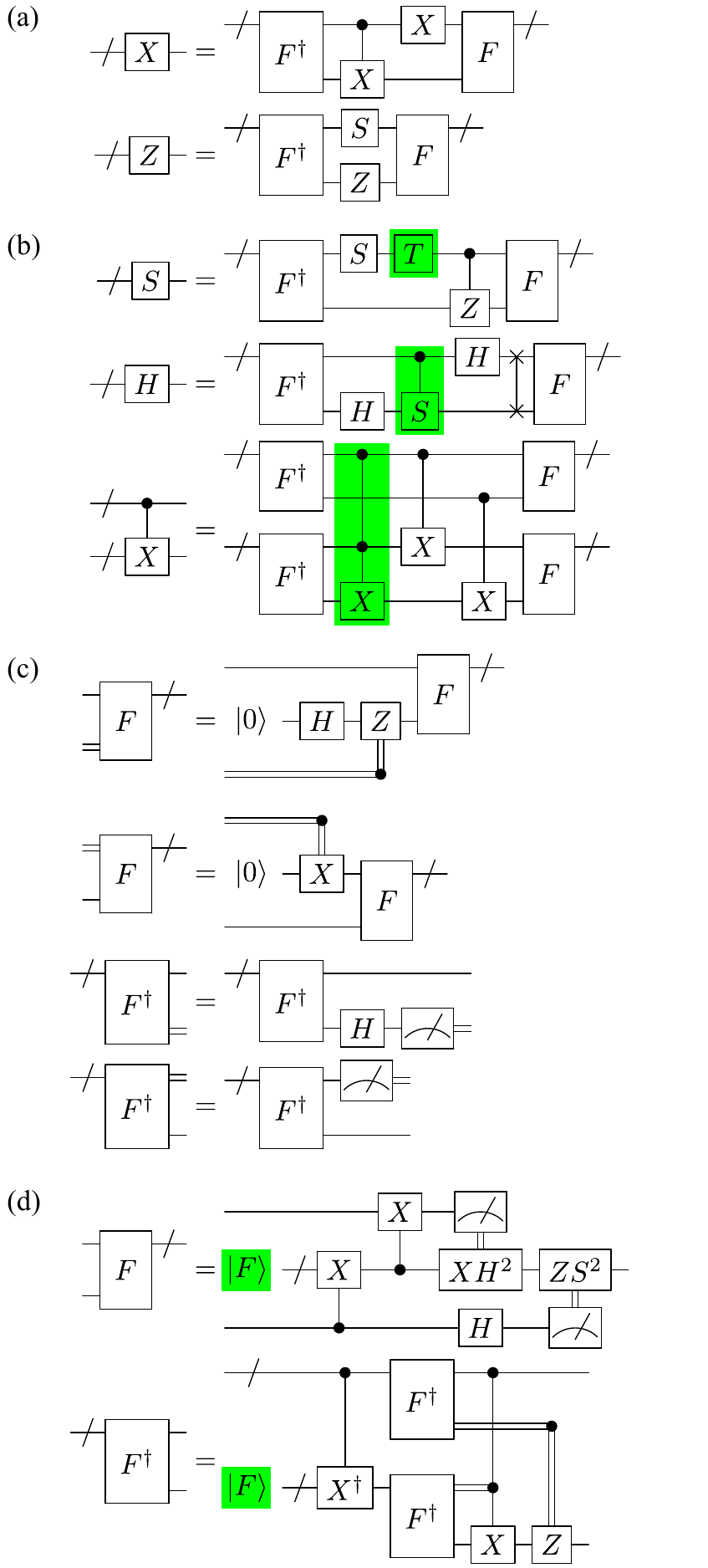}
\caption{\label{fig_fusion} Quantum circuits for qubit fusion ($F$) and qudit fission ($F^\dag$) \cite{qubit_fusion} that convert between a qubit pair and a four-dimensional qudit.
(a) Qudit Pauli gates implemented using only qubit Clifford gates.
(b) Qudit Clifford gates implemented using qubit Clifford gates and several non-Clifford gates (highlighted).
(c) Partial qubit fusion and qudit fission gates with a quantum wire replaced by a classical wire,
 which can be implemented as hybrid stabilizer circuits (not shown).
(d) Full qubit fusion and qudit fission gates implemented as hybrid stabilizer circuits that consume a resource state $|F \rangle$ (highlighted).
}
\end{figure}

Qudits are of limited interest relative to qubits because most experimental efforts are focused on the fabrication of physical qubits.
The Hilbert space of $\lceil \log_2 d \rceil$ qubits is large enough to embed a $d$-dimensional qudit, but it might be difficult to build
 effective qudit Pauli and Clifford gates from qubit gates on the underlying physical qubits.
One example of an embedding is a $d=4$ qudit on a pair of qubits, which was recently studied in terms of quantum circuits \cite{qubit_fusion}.
The qubit fusion gate,
\begin{equation}
 F = \sum_{x=0}^1 \sum_{y=0}^1 |x+2y\rangle \langle x | \otimes \langle y | ,
\end{equation}
 and its Hermitian conjugate, the qudit fission gate $F^\dag$, are used
 to convert between two qubit wires and a qudit wire in circuit diagrams as depicted in Fig. \ref{fig_fusion}.
The combination of $F$ and $F^\dag$ gates with hybrid stabilizer circuits on both qubits and qudits
 is surprisingly useful for universal quantum computation.

At the physical level, $F$ and $F^\dag$ gates exist only as notation and are not actually physical operations.
As shown in Fig. \ref{fig_fusion}, this notation reveals an implementation of qudit Pauli gates in terms of Clifford gates on the underlying qubits.
In contrast, all qudit Clifford gates require a standard non-Clifford gate on one or more qubits.
These gates include the $T$, controlled-$S$, and controlled-controlled-$X$ (Toffoli) gates,
\begin{align}
 T &= \sum_{x=0}^1 \left(\frac{1+i}{\sqrt{2}}\right)^x |x\rangle \langle x | , \\
 CS &= \sum_{x=0}^1 \sum_{y=0}^1 i^{xy} |x\rangle \langle x | \otimes |y\rangle \langle y| , \\
 CCX &= \sum_{x=0}^1 \sum_{y=0}^1 \sum_{z=0}^1 |x\rangle \langle x | \otimes |y\rangle \langle y| \otimes |(z + xy) \bmod 2\rangle \langle z | .
\end{align}
Thus implementations of Pauli and Clifford gates on  physical qudits can be reduced to well-known few-qubit gates.

Hybrid stabilizer circuits merge disjoint stabilizer circuits on qubits and qudits with hybrid Clifford gates between qubits and qudits.
The natural generalization of the $Z$, $S$, $H$, and $CX$ gates that generate the Clifford group is to define controlled-$X$ gates between qubits ($b$) and qudits ($d$) as
\begin{align}
 C_b X_d &= \sum_{x=0}^1 \sum_{y=0}^3 |x\rangle \langle x| \otimes |(y + 2x) \bmod 4\rangle \langle y| , \\
 C_d X_b &= \sum_{x=0}^3 \sum_{y=0}^1 |x\rangle \langle x| \otimes |(y + x) \bmod 2\rangle \langle y| . 
\end{align}
Since $F$ and $F^\dag$ gates transform qudit Clifford gates into qubit non-Clifford gates, they cannot be implemented using hybrid stabilizer circuits.
However, we can still implement partial $F$ and $F^\dag$ gates in Fig. \ref{fig_fusion}c using hybrid stabilizer circuits \cite{qubit_fusion}.

At the logical level, the non-Clifford nature of the $F$ and $F^\dag$ gates facilitates universal quantum computation.
The standard recipe for universality is to insert $T$ gates into qubit stabilizer circuits,
 which are implemented using gate teleportation and resource states.
Alternatively, we can insert $F$ and $F^\dag$ gates into hybrid stabilizer circuits to achieve universality.
This can also be implemented with a qudit nonstabilizer resource state,
\begin{equation}
 | F \rangle = \frac{|0\rangle + |1\rangle}{\sqrt{2}} = F \left( \frac{|0\rangle + |1\rangle}{\sqrt{2}} \otimes |0\rangle \right),
\end{equation}
 and gate teleportation as depicted in Fig. \ref{fig_fusion}d.
$|F\rangle$ is simply an $F$ gate applied to a 2-qubit stabilizer state,
 and it is distillable from noisy $F$ gates \cite{qubit_fusion} analogous to magic state distillation for $T$ gates \cite{magic_state_distillation}.
Qubit fusion offers two possible advantages over magic state distillation.
First, distillation of $|F\rangle$ can take advantage of more general stabilizer codes on both qubits and qudits in the search for more efficient distillation techniques.
Second, it might be possible to directly implement logical $F$ and $F^\dag$ gates as a code conversion between logical qubits and qudits.
In either case, the practical utility of qubit fusion relies on the benefits of logical $F$ and $F^\dag$ gates
 to overcome the cost of hybrid quantum error correction on qubits and qudits.

\begin{figure*}[t]
\includegraphics{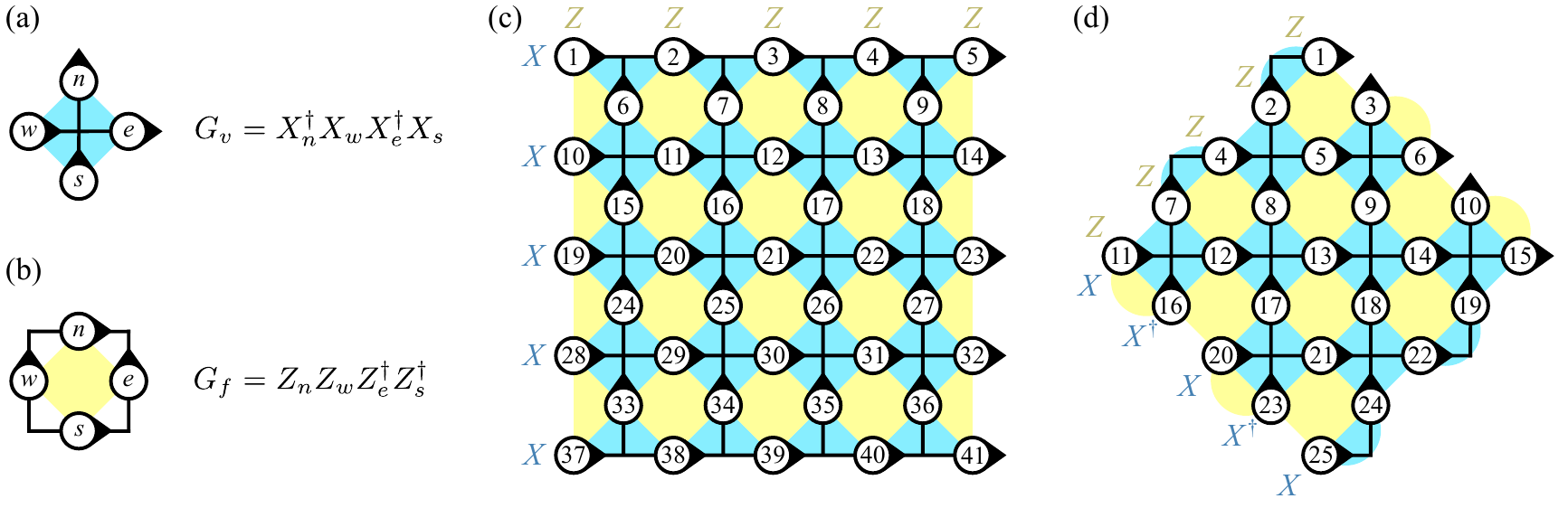}
\caption{\label{fig_stabilizer} $X$-type and $Z$-type generators of the stabilizer group for qudit surface codes.
(a) Vertices denote $X$ or $X^\dag$ on the connected edges, set by direction (in or out).
(b) Faces denote $Z$ or $Z^\dag$ on the surrounding edges, set by orientation (clockwise or counterclockwise).
The complete set of stabilizer generators is defined by a directed graph with qudits on its edges.
Two examples are the distance-5 (c) square and (d) diamond surface codes.
Logical operators $\overline{X}$ and $\overline{Z}$ of these codes are depicted as strings of $X$ and $Z$ operators next to the qudits that they act on.
 }
\end{figure*}

\subsection{Surface codes\label{surface_subsection}}

Surface codes are an important idea in physics that serve as a nexus between topologically protected phases of matter and quantum error correction.
They are also ideally suited to the technological limitations of practical quantum computing
 on a two-dimensional lattice of qubits with physical error rates much larger than the logical error rates necessary for useful computation \cite{resource_estimates}.
We briefly overview surface codes as they have developed historically, from a curiosity in topological physics to a practical design for quantum error correction.

The original topological code construction \cite{quantum_double} considered qudits and qudit operations
 with a general group structure but was restricted to a torus without boundaries.
It was followed by a planar construction with boundaries for qubits \cite{surface_code}, which first defined the modern surface code.
General qudit boundary constructions are a more recent development \cite{double_boundaries}.
We show two examples of distance-5 surface codes in Fig. \ref{fig_stabilizer} specific to the qudit Pauli group in Sec. \ref{qudits}.
The standard notation uses a directed planar graph on a finite surface to define a consistent set of
 qudits (graph edges), $X$-type stabilizer generators (graph vertices), and $Z$-type stabilizer generators (graph faces).
We also use a more recent notation \cite{surface_plaquette} that represents stabilizer generators as shaded tiles with qudits on their vertices.

The surface code construction guarantees that all stabilizer generators commute with each other, 
 with $X$-type strings that join vertex-terminated ``smooth'' boundaries,
 and with $Z$-type strings that join face-terminated ``rough''  boundaries.
There is one string of each type, $\overline{X}$ and $\overline{Z}$, for each example in Fig. \ref{fig_stabilizer}.
For each vertex $v$, there is a generator $G_v$ that detects an error $E_v$,
 which is a $Z$-type string that joins an edge containing $v$ to a rough boundary.
For each face $f$, there is a generator $G_f$ that detects an error $E_f$,
 which is an $X$-type string that joins an edge of $f$ to a smooth boundary.
All of these operator pairs have commutation relations like $X$ and $Z$ in Eqs. (\ref{qudit_X}) and (\ref{qudit_Z})
 and commute with other pairs.
These operators generate the combined Pauli group of all the qudits in the surface code.

The examples in Fig. \ref{fig_stabilizer} generalize to arbitrary distance $D$.
The square contains $2 D^2 - 2 D + 1$ qudits, $D^2 - D$ vertices and faces, and logical strings $\overline{X}$ and $\overline{Z}$ of length $D$.
The diamond contains $D^2$ qudits and also has logical strings with length $D$.
For odd $D$, there are $(D^2 - 1)/2$ vertices and faces.
For even $D$, the symmetry of the boundaries must be lowered, and we consider the case with $D^2/2 - 1$ vertices and $D^2/2$ faces.
The number of qudits minus the number of vertices plus faces is one in each case,
 which enables one logical qudit.
There exists a unitary transformation $U_s$ that transforms each $(X_n , Z_n)$ pair of physical qudit operators to a distinct operator pair from one of the following types:
 $(E_{v}, G_{v})$, $(E_{f}, G_{f})$, or $(\overline{X}, \overline{Z})$.
$U_s$ can be implemented as a depth-$O(D^2)$ stabilizer circuit \cite{quantum_memory}.

The surface code stabilizer generators define a topological phase as the ground-state subspace of a Hamiltonian,
\begin{equation}
 H_s = \sum_{v \in \mathcal{V}} ( 2 - G_v  - G_v^\dag ) + \sum_{f \in \mathcal{F}} (2 - G_f - G_f^\dag),
\end{equation}
 for the set of vertices $\mathcal{V}$ and faces $\mathcal{F}$ of the surface code graph.
By simultaneously diagonalizing the stabilizer generators and $\overline{Z}$,
 we define the $d$ zero-energy ground states of $H_s$ as
\begin{align}
 G_v |\overline{n}\rangle &= |\overline{n}\rangle \ \ \ \forall v \in \mathcal{V} , \label{Vstable} \\
 G_f |\overline{n}\rangle &= |\overline{n}\rangle \ \ \ \forall f \in \mathcal{F} , \label{Fstable} \\
 \overline{Z}  |\overline{n}\rangle &= \omega^n |\overline{n}\rangle , \label{Lstable}
\end{align}
 for $0 \le n \le d-1$.
For quantum computation applications, we would apply $H_s$ to preserve an arbitrary zero-energy state $|\Psi\rangle$.
The 4-spin interactions in $H_s$ are hard to implement directly,
 but $H_s$ can be implemented as the low-energy effective model
 of a Hamiltonian with only 2-spin interactions \cite{honeycomb_model}.
While $H_s$ will have imperfections, its ground-state degeneracy is robust to small pertubations.
However, $|\Psi\rangle$ is not stable with respect to local thermal processes.
Low-energy excited states such as $E_v |\Psi\rangle$ can hop between neighboring vertices at no energy cost
 and return to a zero-energy state as $\overline{Z} |\Psi\rangle$ instead of $|\Psi\rangle$.

\begin{figure}[t]
\includegraphics{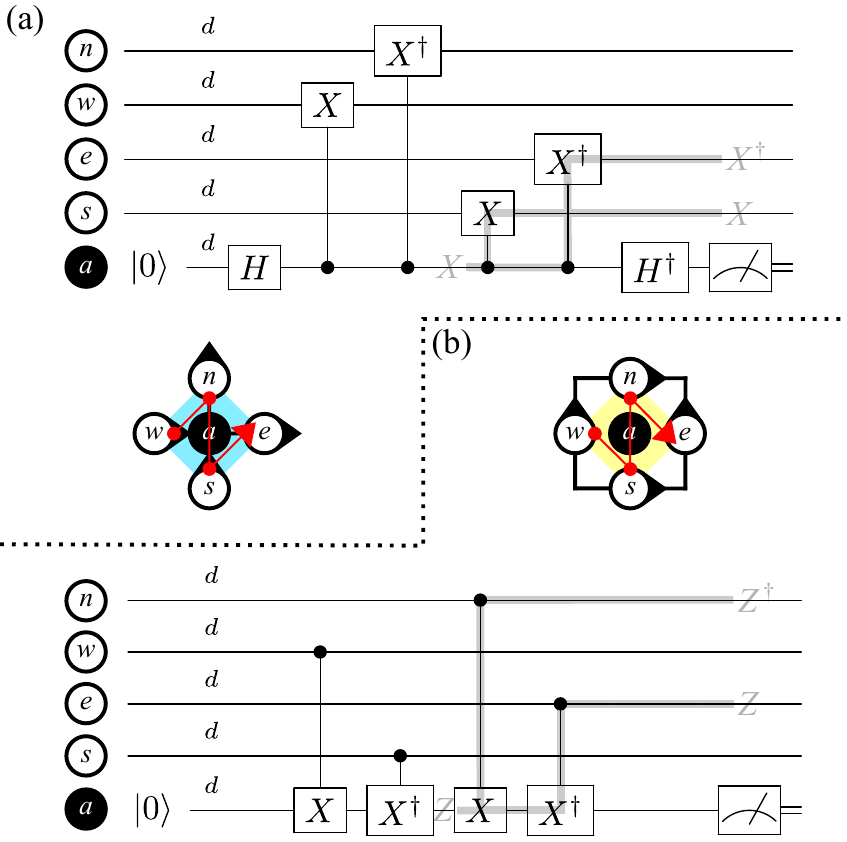}
\caption{\label{fig_measure} Two syndrome measurement circuits to measure (a) $X$-type and (b) $Z$-type stabilizer generators
 of a qudit surface code using an ancilla qudit (black circle), 1-qudit operations, and nearest-neighbor 2-qudit gates.
Stabilizer generators on edges that have fewer qudits simply omit $CX$ gates that have a missing qudit.
These circuits are designed to run simultaneously on all data and ancilla qudits using a qudit ordering depicted by the segmented arrows
 to avoid collisions and misorderings of 2-qudit gates.
They can produce ``hook'' errors depicted in gray, whereby an ancilla error between the second and third $CX$ gate spreads to two data qudits.
With these qudit orderings adapted from recent work on qubits \cite{measurement_circuit},
 we only spread hook errors perpendicular to the direction of logical Pauli strings on the diamond surface code in Fig. \ref{fig_stabilizer}d
 to limit their contribution to logical errors.
}
\end{figure}

The thermal instability of Hamiltonian-based passive error correction is resolved by circuit-based active error correction \cite{quantum_memory}.
Stabilizer generators are repeatedly measured, and their ``syndrome'' measurement outcomes enable the identification and correction of physical qudit errors.
With the ``data'' qudits shown in Fig. \ref{fig_stabilizer} interleaved with ``ancilla'' qudits, a stabilizer generator can be measured using
 only 1-qudit operations and nearest-neighbor 2-qudit gates as shown in Fig. \ref{fig_measure}.
Assuming that physical qudit errors are weakly correlated in both space and time,
 the occurrence of logical Pauli errors with the form $\overline{X}^m \overline{Z}^n$ can be inferred statistically from a history of syndrome measurements.
Decoding logical errors is straightforward for concatenated low-distance codes \cite{decode_concatenate}.
Optimal decoding of a high-distance surface code is more difficult \cite{decode_surface}, but efficient heuristic decoders perform well in practice
  such as minimum weight perfect matching \cite{decode_matching} for qubits and renormalization group decoders for qudits \cite{decode_RG}.
With the possibility of errors in syndrome measurements, decoding requires the past $O(D)$ rounds of measurements to decode a distance-$D$ code.
Even if there are no errors on any data qudits, a majority vote for each syndrome measurement is necessary to suppress measurement errors
 from a high physical rate $\epsilon$ to a low logical rate $\epsilon^{O(D)}$.

Once successful error correction has been established, the next important component of quantum computation is logical state preparation and measurement.
We assume the ability to prepare physical qudits in their zero state, which produces the joint physical zero state $|0\rangle$ defined by stabilizer conditions
\begin{equation}
 Z_n |0\rangle = |0\rangle .
\end{equation}
Our goal is to prepare the logical zero state $|\overline{0}\rangle$ defined by Eqs. (\ref{Vstable}), (\ref{Fstable}), and (\ref{Lstable}).
In principle, we can transform between these two states and stabilizer conditions by applying $U_s$ to $|0\rangle$ as $|\overline{0}\rangle = U_s |0\rangle$.
In practice, $U_s$ is a high-depth stabilizer circuit that is avoided by exploiting active error correction.
We first prepare $|0\rangle$ and then measure the stabilizer generators of the surface code.
In the absence of errors, Eqs. (\ref{Fstable}) and (\ref{Lstable}) are satisfied by $|0\rangle$,
 and all measurement outcomes for $G_f$ and $\overline{Z}$ are one.
Eq. (\ref{Vstable}) is not satisfied by $|0\rangle$, and the measurement outcomes for $G_v$ are random.
However, $|0\rangle$ is projected into a simultaneous eigenstate of all $G_v$,
 which can be transformed to $|\overline{0}\rangle$ by applying $E_v$ to switch to the $\omega^0$ eigenstate of each $G_v$.
To measure $\overline{Z}$ destructively in the absence of errors, we measure each $Z_n$
 and reconstruct the $\overline{Z}$ measurement outcome from a product of $Z_n$ measurement outcomes.
These methods for logical state preparation and measurement are sensitive to physical qudit errors
 and can be incorporated into active error correction as boundary conditions for decoding \cite{quantum_memory}.

To fully enable quantum computation, we must implement a universal set of logical gates on a surface code.
In the qubit case, this has been achieved \cite{surface_operations} by combining a limited set of logical Clifford gates
 with resource state distillation and gate teleportation.
This framework has not yet been generalized to qudits, but existing results indirectly suggest that it is possible.
The transversal Clifford gates of qubit color codes \cite{color_code} have a straightforward generalization to qudit color codes \cite{qudit_color}.
The use of magic state distillation to prepare a qubit non-Clifford gate \cite{magic_state_distillation} has also been generalized to qudits \cite{qudit_magic_states},
 but only for prime dimension $d$.
The efficiency of a distillation protocol is roughly determined by the ratio of input states to output states.
For quadratic error reduction, the best practical qubit protocol is 5-to-1 \cite{qubit_distillation}
 and the best known qudit protocol is $(d-1)$-to-1 for prime $d \ge 5$ \cite{qudit_magic_states}.
Thus, $d=5$ qudits are capable of more efficient magic state distillation than qubits until the regime of
 asymptotic 3-to-1 \cite{asymptotic_distillation} and 2-to-1 \cite{asymptotic_distillation2} multi-qubit protocols.
Magic state distillation is not yet developed for general $d$.

\subsection{Code symmetries\label{Steane_subsection}}

\begin{figure}[t]
\includegraphics{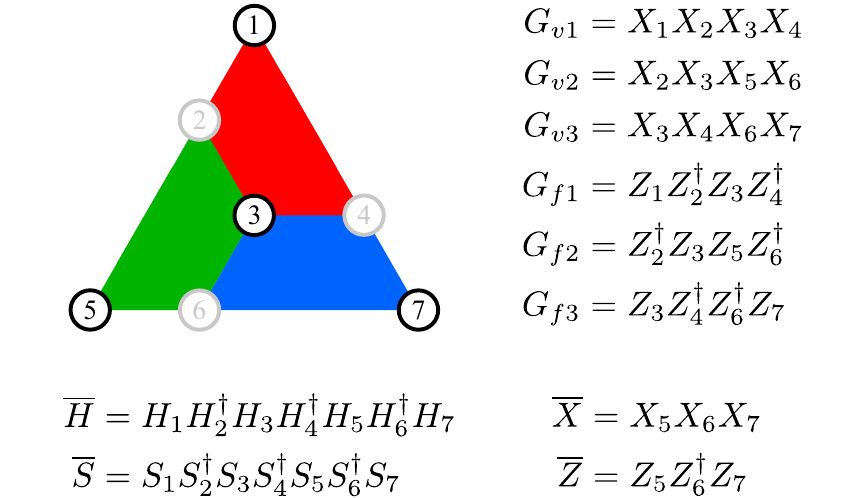}
\caption{\label{fig_Steane} Distance-3 qudit code that generalizes the Steane code \cite{qudit_color}.
Data qudits are partitioned into two types (black and gray) as vertices of a bipartite graph with edges corresponding to edges of the shaded tiles.
Each tile defines one $X$-type and one $Z$-type stabilizer generator on its qudit vertices.
We label the generators with a notation similar to Fig. \ref{fig_stabilizer}.
The logical Pauli and Clifford gates are ``star transversal'' and decompose into single-qudit gates that depend on qudit type.
}
\end{figure}

Certain symmetries of quantum error correcting codes can simplify the implementation of logical gates.
We consider the case of logical Clifford gates on the qudit Steane code,
 where there is high symmetry and simple gate implementations.
The development of folded surface codes in Sec. \ref{fold} follows from
 this example by identifying the reduced symmetries of these codes and adapting the gate implementations from the Steane code.
We are guided by a formal equivalence between folded surface codes and color codes for qubits \cite{fold_surface}.
The distance-3 color code is the Steane code, which is depicted in Fig. \ref{fig_Steane}.

Among logical gates, Clifford gates are particularly simple because they can be implemented as unitary stabilizer circuits.
Just as conjugation by physical Clifford operators transforms between physical Pauli operators as in Eq. (\ref{conjugation}),
 we expect that conjugation by logical Clifford operators transforms between logical Pauli operators in some sense.
However, we only care about the action of logical Pauli operators on logical states and not the operators themselves,
 which expands the concept of a logical Clifford operator relative to the physical case.
We can multiply a logical Pauli operator $\overline{P}$ by a product of stabilizer generators
 and its action on a logical state $|\Psi\rangle$ will not change because of Eqs. (\ref{Vstable}) and (\ref{Fstable}).
If we define a stabilizer group $\mathcal{S}$ from the products of stabilizer generators,
 then the products of generators with $\overline{P}$ form the coset $\overline{P} \mathcal{S}$.
Conjugation by any logical Clifford operator $\overline{U}$ must then transform elements of $\overline{P} \mathcal{S}$ to
 $\overline{Q} \mathcal{S}$ for some other logical Pauli operator $\overline{Q}$ to have
 the action $\overline{U} \overline{P} \overline{U}^\dag |\Psi\rangle = \overline{Q} |\Psi\rangle$.
These conditions reduce to
\begin{align}
 \overline{U} \overline{P} \overline{U}^\dag & \in \mathcal{L} \mathcal{S} , \label{logical_Clifford1} \\
 \overline{U} G \overline{U}^\dag & \in \mathcal{S} \label{logical_Clifford2}
\end{align}
 for all logical Pauli generators $\overline{P}$ of the logical Pauli group $\mathcal{L}$ and
 for all stabilizer generators $G$ of the stabilizer group $\mathcal{S}$.

The qubit Steane code has self-dual symmetry, which pairs
 every $X$-type stabilizer generator with a $Z$-type generator
 that results from a uniform $X \leftrightarrow Z$ substitution.
The logical Pauli generators $\overline{X}_L$ and $\overline{Z}_L$ are similarly paired.
For qudits, this is complicated by ``star bipartition'' structure \cite{qudit_color}, where qudits
 are separated into two sets as vertices of a bipartite graph.
A pair of generators is defined by $X \leftrightarrow Z$ substitution on one set of qudits and $X \leftrightarrow Z^\dag$ on the other.
If $X \leftrightarrow Z$ substitution is used for all qudits, commutation between $X$-type and $Z$-type stabilizer generators produces an $\omega^4$ phase factor
 on the same tile and $\omega^2$ between tiles.
The star bipartition produces pairs of $\omega$ and $\omega^{-1}$ to guarantee that stabilizers commute.

The self-dual symmetry of the qudit Steane code facilitates transversal logical Clifford gates, which are shown in Fig. \ref{fig_Steane}.
The $\overline{H}$ gate satisfies Eq. (\ref{logical_Clifford1}) by exchanging $\overline{X}$ and $\overline{Z}$,
\begin{align}
 \overline{H} \overline{X}\overline{H}^\dag & = \overline{Z} \notag \\
 \overline{H} \overline{Z} \overline{H}^\dag & = \overline{X}^\dag , \label{H_logical}
\end{align}
 and satisfies Eq. (\ref{logical_Clifford2}) by exchanging $G_v$ and $G_f$ on a tile,
\begin{align}
 \overline{H} G_v \overline{H}^\dag & = G_f \notag \\
 \overline{H} G_f \overline{H}^\dag & = G_v^\dag . \label{H_stable}
\end{align}
Similarly, the $\overline{S}$ gate cancels pairs of $\omega^{-1/2}$ and $\omega^{1/2}$ phases to
 produce the correct phase for its logical action in Eq. (\ref{logical_Clifford1}),
\begin{align}
 \overline{S} \overline{X} \overline{S}^\dag & = - \omega^{-1/2} \overline{X} \overline{Z} \notag \\
 \overline{S} \overline{Z} \overline{S}^\dag & = \overline{Z} , \label{S_logical}
\end{align}
 and mix $G_v$ and $G_f$ on a tile with no phase as in Eq. (\ref{logical_Clifford2}),
\begin{align}
 \overline{S} G_v \overline{S}^\dag & = G_v G_f \notag \\
 \overline{S} G_f \overline{S}^\dag & = G_f . \label{S_stable}
\end{align}
Without the star bipartition structure, the conjugation of $G_v$ by $\overline{S}$ would produce an $\omega^2$ phase
 from a conjugation of four $X$ by four $S$.
This multiplicity of four in the weight of $X$-type stabilizers is referred to as ``doubly even'' code structure \cite{CSS_codes}.
Doubly even self-dual qubit codes have strongly transversal $S$ gates since $\omega^2 = 1$,
 but this does not generalize to qudits.

For a pair of Steane codes, we can construct logical 2-qudit Clifford gates that do not need complete self-dual symmetry.
We add a label on all operators to distinguish between codes.
The first gate is a ``mirror'' gate $\overline{M}$ that combines $\overline{H}_1$, $\overline{H}_2$, and a strongly transversal qudit swap gate between the codes.
It acts like $\overline{H}$ in Eqs. (\ref{H_logical}) and (\ref{H_stable}),
\begin{align}
 \overline{M} \overline{X}_1 \overline{M}^\dag & = \overline{Z}_2 \notag \\
 \overline{M} \overline{Z}_1 \overline{M}^\dag & = \overline{X}_2^\dag  \notag \\
 \overline{M} G_{v,1} \overline{M}^\dag & = G_{f,2} \notag \\
 \overline{M} G_{f,1} \overline{M}^\dag & = G_{v,2}^\dag . \label{M_conjugate}
\end{align}
The second gate is the star transversal controlled-$Z$ gate $\overline{CZ}$ from $\overline{Z}$ in Fig. \ref{fig_Steane}.
It acts like $\overline{S}$ in Eqs. (\ref{S_logical}) and (\ref{S_stable}),
\begin{align}
 \overline{CZ} \, \overline{X}_1 \overline{CZ}^\dag & = \overline{X}_1 \overline{Z}_2 \notag \\
 \overline{CZ} \, \overline{Z}_1 \overline{CZ}^\dag & = \overline{Z}_1 \notag \\
 \overline{CZ} G_{v,1} \overline{CZ}^\dag & = G_{v,1} G_{f,2} \notag \\
 \overline{CZ} G_{f,1} \overline{CZ}^\dag & =G_{f,1} . \label{CZ_conjugate}
\end{align}
Eqs. (\ref{M_conjugate}) and (\ref{CZ_conjugate}) pair an $X$-type operator on one code with a $Z$-type operator on the other,
 rather than in the same code.

\section{Folded surface codes\label{fold}}

The concept of a folded surface code originates from recent studies of the relationship between surface and color codes of finite extent \cite{fold_surface}.
A logical qubit encoded in a color code can be re-encoded in a folded surface code through a local unitary transformation involving ancilla qubits.
Limitations on color code construction and the transformation process restrict the types of surface codes that can be produced.
For example, the standard surface code in Sec. \ref{surface_subsection} with four edges per vertex and per face is unattainable from known color codes.
We take a complementary approach of folding standard surface codes to gain some of the properties of color codes rather than fully transforming to a color code.
This inverse approach has fewer constraints, and it can be adapted to construct folded versions of both the square and diamond surface codes in Fig. \ref{fig_stabilizer}.

The property of color codes that we seek to adapt to folded surface codes is the transversality of logical Clifford gates.
As in the Steane code example from Sec. \ref{Steane_subsection}, qudit color codes have transversal Clifford gates because of self-dual symmetry and star bipartition structure.
Each tile of a color code such as in Fig. \ref{fig_Steane} corresponds to both an $X$-type and $Z$-type stabilizer generator,
 which are mixed when conjugated by the logical $H$ or $S$ gate as in Eqs. (\ref{H_stable}) and (\ref{S_stable}).
Each tile of a surface code such as in Fig. \ref{fig_stabilizer} only corresponds to one type, either $X$ or $Z$, of stabilizer generator,
 and any star transversal $H$ and $S$ gates do not preserve the stabilizer group.
The premise of folding a surface code is to pair an $X$-type generator in one layer with a $Z$-type generator in the other layer.
For qudits along the fold, there will be partial overlap of $X$-type and $Z$-type generators.
Logical Pauli operators must also pair, an $X$-type string on one layer with a $Z$-type string on the other, and overlap on a fold.
We refer to this relaxation of self-dual symmetry with qudits partitioned into two layers and a fold as ``mirror dual.''

The star bipartition structure of qudit color codes naturally generalizes to qudit surface codes with mirror-dual symmetry.
When the directed graph that defines a surface code is folded,
 the subgraph on one layer is dual to the subgraph on the other layer.
The fold qudits and overlapping pairs of qudits between layers are grouped into two types
 based on the direction of the cross product between the overlapping edges of the subgraphs.
This version of star bipartition structure is more general than the color code version.
The graph with qudits as vertices and tile edges as edges does not need to be bipartite
 and no longer defines the qudit type.
Also, $X$-type stabilizer generators of a surface code contain both $X$ and $X^\dag$ rather than only $X$.

\begin{figure}[t]
\includegraphics{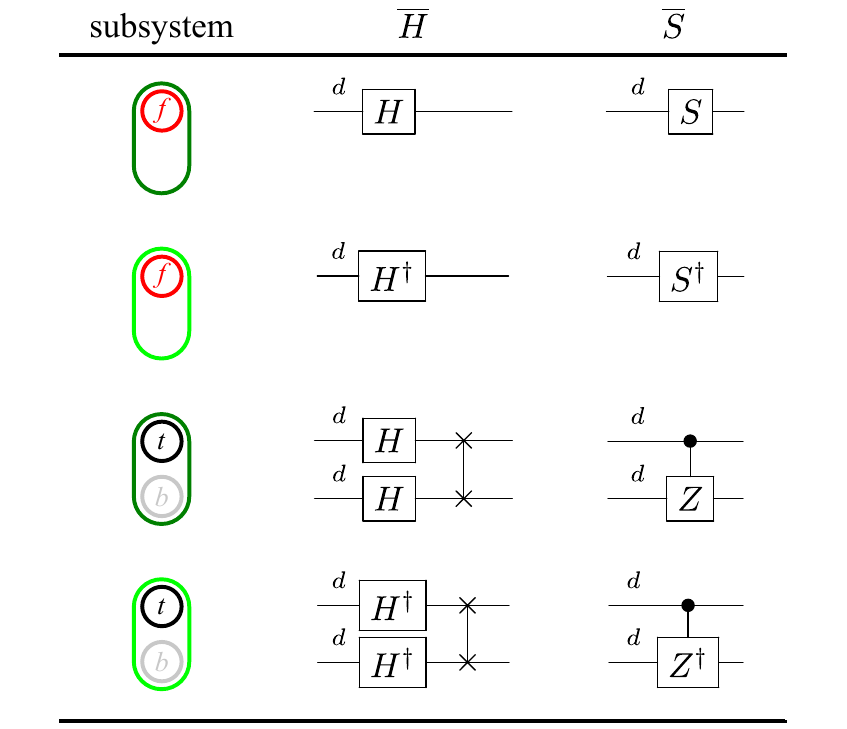}
\caption{\label{fig_logical} Transversal implementation of the $\overline{H}$ and $\overline{S}$ logical Clifford gates
 for folded surface codes organized into circuits on four distinct subsystems.
These gates act differently on qudits along the fold (``$f$'') and pairs of qudits on the top (``$t$'') and bottom (``$b$'') layers.
They also act differently on the two types of qudits (in light and dark stadiums) defined by the star bipartition.
These circuits do not depend on qudit orientation in subsystems, which can be altered by other operations.
}
\end{figure}

We can build logical Clifford gates on a qudit surface code with mirror-dual symmetry by using concepts from Sec. \ref{Steane_subsection}.
The $\overline{H}$ gate combines Eqs. (\ref{H_logical}), (\ref{H_stable}), and (\ref{M_conjugate}) by performing $H$ and $H^\dag$ on fold qudits
 and a mirror gate on paired qudits.
The $\overline{S}$ gate combines Eqs. (\ref{S_logical}), (\ref{S_stable}), and (\ref{CZ_conjugate}) by performing $S$ and $S^\dag$ on fold qudits
 and $CZ$ and $CZ^\dag$ on paired qudits.
The general implementation of these gates is shown in Fig. \ref{fig_logical}.
The star transversality of logical Clifford gates on qudit color codes is relaxed,
 and gates can spread errors between layers of the folded surface code within 2-qudit transversal subsystems.
This spreading of errors must be considered when studying the fault tolerance of a folded surface code.

Conceptually, a folded surface code has two layers of qudits stacked on top of each other.
Practically, we can implement a folded code on a planar grid of qudits by partitioning it into a grid of qudit pairs as shown in Fig. \ref{fig_square}.
Data qudit pairs and ancilla qudit pairs are arranged in a checkerboard pattern.
The top layer of a surface code is depicted with black qudits on the edges of a black graph,
 and the bottom layer is depicted with gray qudits on the edges of a gray graph.
All top and bottom qudits are paired in a mirror-dual code, and qudits along a fold are paired with additional ancilla qudits.
For simplicity, we do not display shaded tiles or graph edge directions as in Fig. \ref{fig_stabilizer}.
We use a convention for the top-layer graph that vertical edges are directed up and horizontal edges are directed right.
Edge directions of the bottom-layer graph are defined indirectly by the direction of their cross product with overlapping directed edges of the top-layer graph.
Dark stadiums surrounding qudit pairs denote cross products pointing out of the plane and light stadiums denote cross products pointing into the plane.

With the same qudit partitioning as in Fig. \ref{fig_square}, we can stack two unfolded surface codes.
This data qudit layout enables a strongly transversal $CX$ gate.
Every Calderbank-Steane-Shor (CSS) code possesses a strongly transversal $CX$ gate \cite{CSS_codes},
 but geometric restrictions often limit its practical implementation.
For example, conventional planar surface codes implement a $CX$ gate with either defect braiding \cite{surface_operations} or lattice surgery \cite{surface_surgery}.
While a transversal $CX$ gate requires a shallower circuit than these alternatives, $O(1)$ versus $O(D)$,
 the stacked surface codes that enable it need deeper circuits for stabilizer measurement as discussed in Sec. \ref{folded_QEC}.
Also, a transversal $CX$ gate spreads errors between stacked codes, which complicates decoding.

With two virtual surface code layers, we have access to the multi-qudit logical Clifford group.
A physical qudit grid tiled with folded surface codes enables transversal implementation of the single-qudit logical Clifford group.
We unfold and stack pairs of surface codes to enable strongly transversal $CX$ gates that generate the multi-qudit logical Clifford group.
This plan requires folded forms of the square and diamond surface code families in Fig. \ref{fig_stabilizer}
 with folding and unfolding operations.
The stacking of unfolded codes requires logical qudit movement.
We use lattice surgery \cite{surface_surgery} to expand a logical qudit over both
 its initial and final locations and then contract it to fit within its final location while preserving the code distance.
With all Clifford gates transversal, the computational bottleneck is the movement of logical quantum information.

\subsection{Folding the square}

\begin{figure}[!t]
\includegraphics{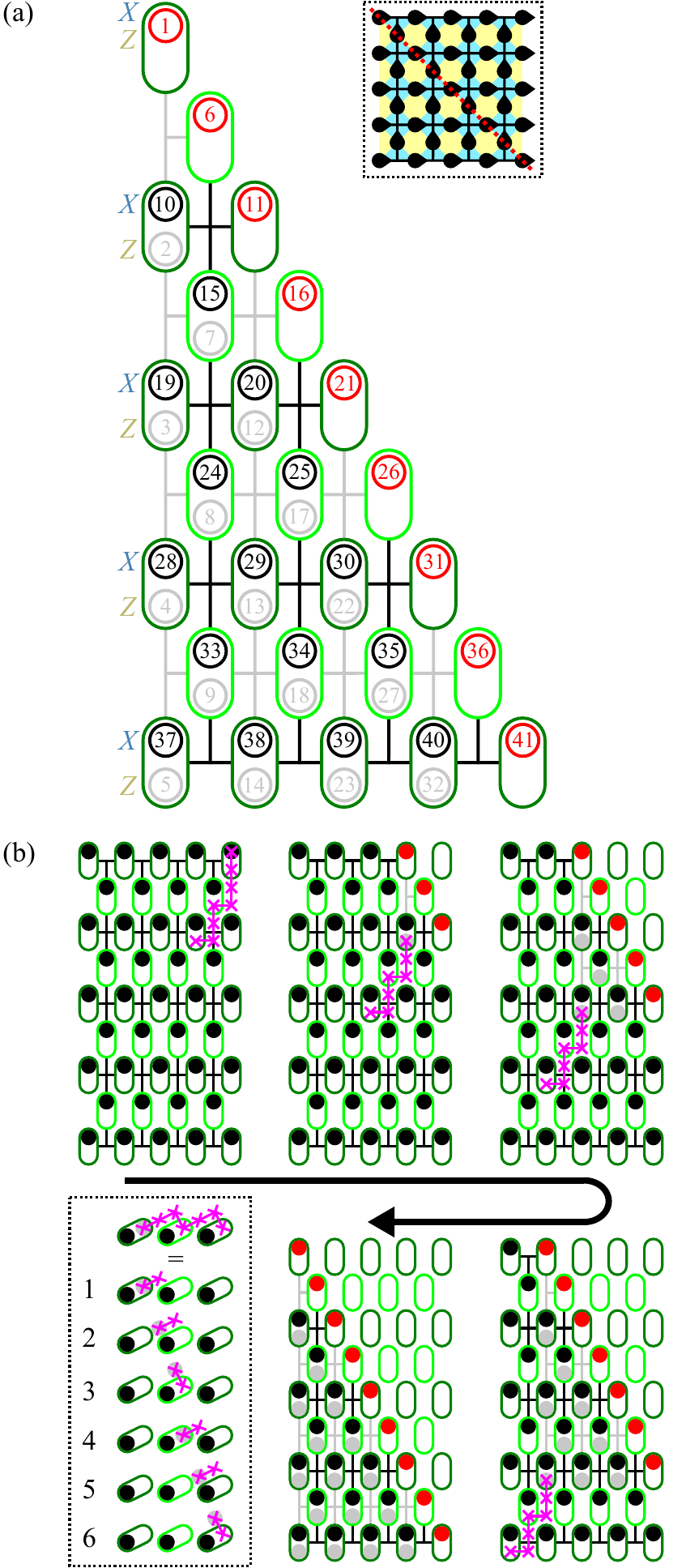}
\caption{\label{fig_square} Distance-5 version of the folded square surface code.
The planar layout of the folded code (a) pairs each qudit on the top layer (black) with a qudit on the bottom layer (gray)
 but qudits along the fold are unpaired.
The folding operation (b) moves qudits with local swaps through a sequence of partially folded surface codes.
As an illustrative example, we show the swap path for the top-right qudit decomposed into 6-swap segments defined in the inset.
}
\end{figure}

To impose mirror-dual symmetry on a square surface code as in Fig. \ref{fig_stabilizer}c, we fold it along a diagonal as shown in Fig. \ref{fig_square}a.
The spatial layout of the code has changed, but its underlying stabilizer structure remains unchanged.
The same transversal $\overline{H}$ and $\overline{S}$ gates are valid for both an unfolded and folded code,
 but the nearest-neighbor 2-qudit gates on a folded code map to
 long-range 2-qudit gates on the unfolded code.

On a virtual 2-layer lattice of qudits, we can fold a square surface code on one layer if the other layer is not being used to encode a logical qudit.
We decompose folding operations into a sequence of local swap operations between data qudits and ancilla qudits.
In Fig. \ref{fig_square}b, we visualize the swap path for the top-right data qudit.
As the longest swap path, it sets the circuit depth of a folding operation to $6 D - 5$ for a distance-$D$ square surface code.
Unlike the transversal Clifford gates, the depth of folding operations increases with code distance.

The folding operation is a logical identity, but it physically deforms the spatial layout of a surface code.
After every step of the folding operation, there is the same surface code with a different layout.
To prevent a build-up of errors during $O(D)$ swap operations, we can perform one round of quantum error correction
 on these partially folded surface codes.
As long as all data qudits are in a 2-qudit block intended for data qudits,
 then the syndrome measurement circuits are similar to those discussed in Sec. \ref{folded_QEC}.
We can reduce the number of swaps per data qudit per round of error correction to three.

This is the first of many instances in this paper where swap operations are prevalent.
For convenience, we consider swaps to be primitive physical operations.
If swaps are implemented using other primitive gates such as $H$ and $CX$,
 then they will have a significantly higher error rate than other primitives.
If swaps are distinct non-entangling physical operations
 such as shuttling operations on ions \cite{ion_shuttling} or electrons \cite{electron_shuttling} that host a physical qudit,
 then they may have an error rate that is more comparable to other primitives.
Folding operations use swaps between data qudits and inactive ancilla qudits,
 which can be achieved with standard teleportation circuits
 that only require one $CX$ gate but measure the ancilla \cite{textbook}.
The best choice of swap will depend on the details of a physical implementation
 of qudits and the primitive gates that are available.

\subsection{Extending the diamond}

\begin{figure}[!t]
\includegraphics{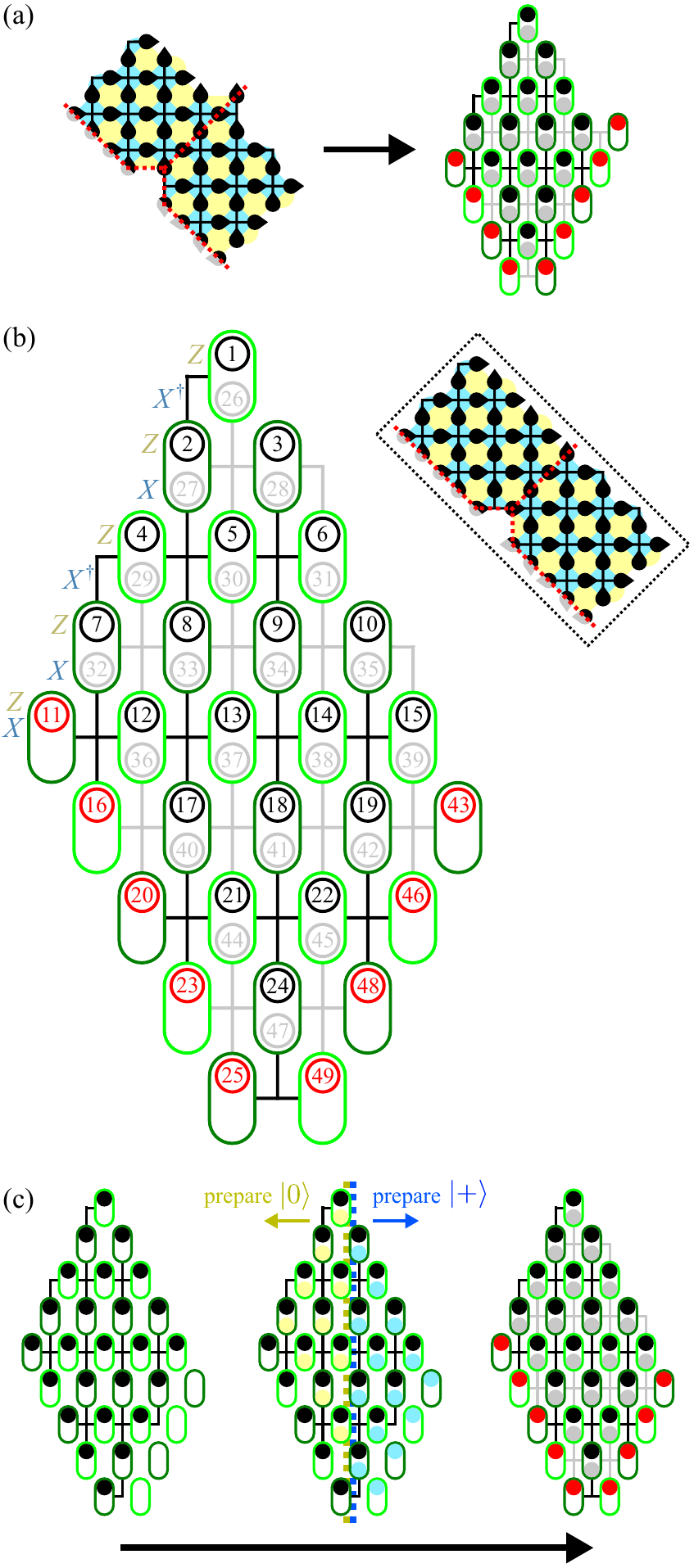}
\caption{\label{fig_cone} Distance-5 version of the folded cone surface codes.
The most efficient construction (a) has a distorted edge to guarantee that stabilizer generators on the edge and tip of the cone commute.
A less efficient construction (b) corrects the edge distortion and generates a stabilizer group that can be reduced to the smaller stabilizer group of a distance-5 diamond surface code.
This compatibility enables code conversion between a diamond and cone (c) by preparing each qudit of one layer ($|+\rangle = H |0\rangle$) and then switching between codes.
}
\end{figure}

Because the diamond surface code has chiral edges, it is not possible to impose mirror-dual symmetry by folding.
Instead, we can create a mirror-dual copy of the diamond surface code and connect two edges to create one logical qudit.
Effectively, this a surface code on a cone that is folded and flattened.
Two distance-5 examples of extended diamond surface codes with mirror-dual symmetry are shown in Fig. \ref{fig_cone}.

The minimal extension of the diamond surface code into a cone is shown in Fig. \ref{fig_cone}a.
In general, it increases the number of data qudits in a distance-$D$ code from $D^2$ to $2 D^2 - 2D + 1$,
 which is equivalent to the square surface code.
Unfortunately, the cone and diamond do not have the same spatial footprint because of a distorted edge,
 which complicates the packing of multiple logical qudits into a large grid of physical qudits.
A more significant problem is that the stabilizer groups of a cone and diamond do not commute,
 which complicates conversion between the two codes.
We can convert while preserving code distance using code deformation \cite{code_deformation},
 but a simpler approach such as lattice surgery \cite{surface_surgery} is desirable.

A less efficient extension of the diamond surface code into a cone is shown in Fig. \ref{fig_cone}b.
It uses $2 D^2 - 1$ physical qudits to encode a logical qudit with distance $D$.
Stabilizer generators of the diamond can all be extended into stabilizer generators of the cone,
 which enables simple code conversion between a diamond and cone using lattice surgery.
These diamonds and cones have different spatial footprints, but we can pack logical qudits
 to merge subsystems containing one fold qudit between neighboring folded codes.
The two cones in Fig. \ref{fig_cone} then have different code efficiencies but the same packing efficiency.

The basic idea of lattice surgery \cite{surface_surgery} is to switch between distance-$D$ codes instantaneously
 and apply $D$ rounds of error correction for fault tolerance.
The conversion from a diamond to a cone is shown in Fig. \ref{fig_cone}c.
Physical qudits are prepared in specific states to enable deterministic measurement outcomes
 for approximately half of the new stabilizer generators in the absence of errors.
The new stabilizer generators with random measurement outcomes cannot detect errors when we switch codes,
 and we choose them to prevent undetected errors from reducing the effective code distance.
To convert from a cone to a diamond,
 we measure excess physical qudits of the cone in the basis that they were prepared in.
With this information, we can relate the syndrome measurements between the cone and diamond
 to detect errors at their code boundary.

\subsection{Syndrome measurements\label{folded_QEC}}

Just as with a color code, a folded surface code simplifies logical Clifford gates
 but complicates syndrome measurement circuits relative to a conventional surface code.
However, we can use the arbitrary orientation of qudit pairs in transversal subsystems to simplify syndrome measurements
 by allowing circuits to change qudit orientation.
This requires additional bookkeeping on the state of a logical qudit.
We discuss a few detailed examples of syndrome measurement circuits
 and then overview the general features of the error correction cycle.

\begin{figure}[!t]
\includegraphics{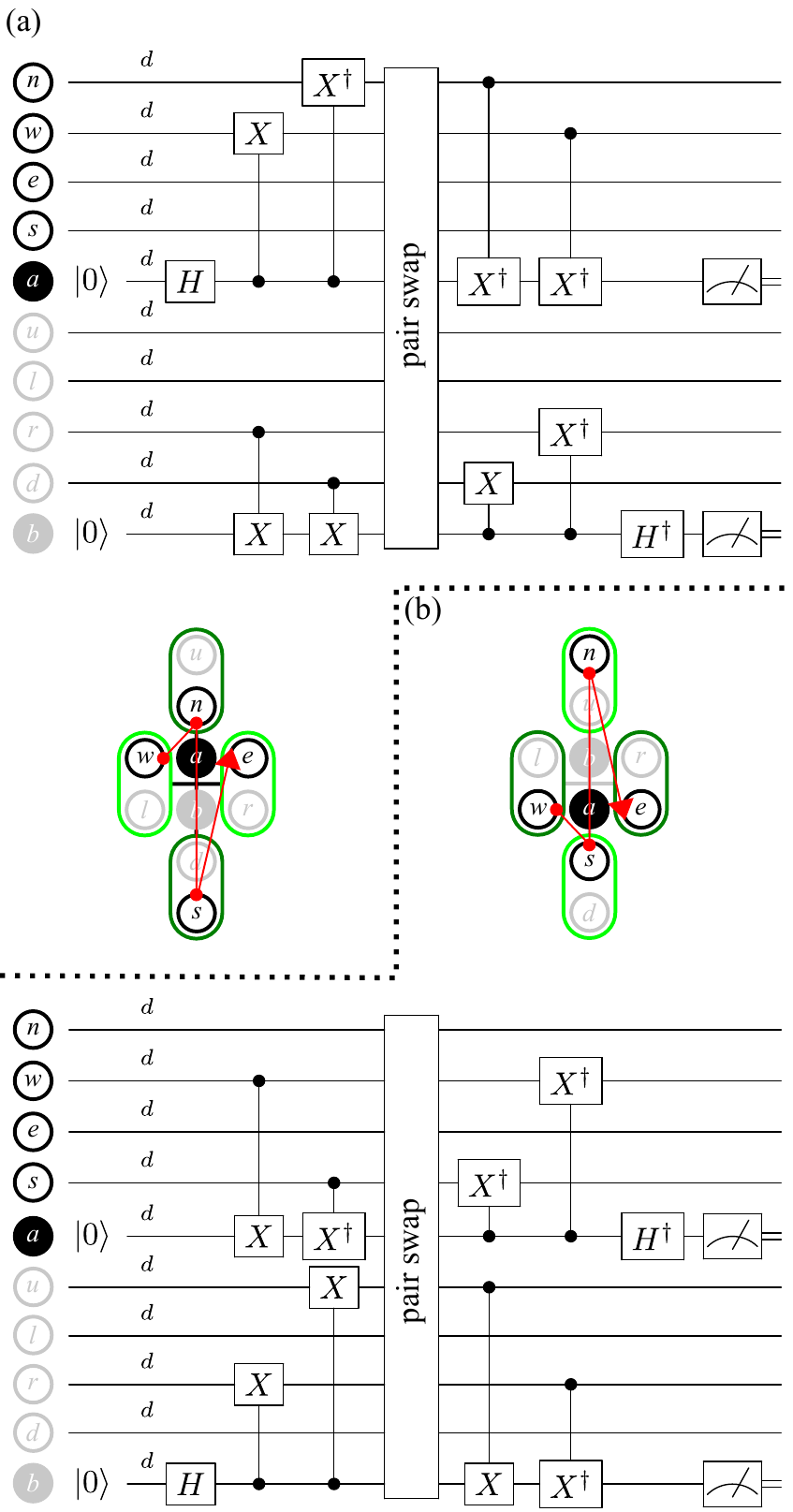}
\caption{\label{fig_foldQEC} Two syndrome measurement circuits to measure stabilizer generator pairs
 of a folded qudit surface code that are adapted from the circuits in Fig. \ref{fig_measure}.
Every $X$-type stabilizer generator on one layer is paired with a $Z$-type stabilizer generator on the other layer,
 either with the $X$-type generator on the (a) top or (b) bottom layer.
A single global pairwise swap operation is needed to implement these circuits using only nearest-neighbor 2-qudit gates. 
From input to output, the data qudits are swapped between the top and bottom layers.
These circuits can be adapted to an alternate data qudit layout with top and bottom layers switched by reversing the time order of $CX$ gates.
}
\end{figure}

If we can implement a folded surface code on two layers of physical qudits,
 then the syndrome measurement circuits are identical to a conventional surface code as in Fig. \ref{fig_measure}.
With a planar layout, these circuits require 2-qudit gates on pairs of qudits that are not nearest neighbors.
To reduce the circuits to only nearest-neighbor 2-qudit gates, we need additional swap operations.
We can limit this overhead to one round of swap operations if we alternate the orientations of top and bottom qudits between rows
 as shown for two examples in Fig. \ref{fig_foldQEC}.
A syndrome measurement circuit changes the orientation of all qudit pairs,
 but requires a specific initial orientation to work correctly.
The syndrome measurement circuit for the alternate orientation is the original circuit with the $CX$ gates applied in the reverse order.
The error correction cycle switches between applying $CX$ gates from left to right and from right to left.

The simultaneous measurement of stabilizer generators is a nontrivial scheduling problem.
The circuits in Fig. \ref{fig_measure} function correctly in the interior of a folded surface code
 but must be altered to function near a fold.
Because overlapping stabilizer generators on the top and bottom layers interact with a qudit along a fold
 on opposite sides of the pairwise swap step, the corresponding ancilla qudits can interact with a fold qudit
 if the fold qudit swaps with the additional ancilla qudit during the pairwise swap step.
All relevant ancilla qudits are able to perform a $CX$ gate with the fold qudit
 without increasing the depth of the measurement circuits.
However, the $X$-type and $Z$-type stabilizer generators that overlap along a fold have an
 incorrectly ordered $CX$ gate in their measurement circuit that induces an extra $CX$ gate between their corresponding ancilla qudits \cite{surface_operations}.
Because these ancilla qudits are nearest neighbors, a local $CX^\dag$ gate can remove these unwanted $CX$ gates.

For a gate-based error model that accounts for the spread of Pauli errors within syndrome measurement circuits,
 increased circuit depth and complexity results in increased diversity and rates of physical errors.
The unfolded syndrome measurement circuits in Fig. \ref{fig_measure} produce ``hook'' errors,
 which are spatially correlated pairs of Pauli errors.
The folded circuits in Fig. \ref{fig_foldQEC} additionally spread errors between layers during the pairwise swap operation.
However, these swaps act within transversal subsystems and do not spread errors between subsystems.
To simplify decoding, we can consider a folded surface code of qudits to be an unfolded surface code of qudit pairs.
When an error spreads between layers, we can interpret it as a change of error type rather than a change in error location.

\subsection{Resource estimates\label{overhead}}

For folded surface codes to become useful in practice, they must be able to reduce the resource requirements of quantum computation in some parameter regime.
We consider the well understood case of qubits and compare with known resource estimates of unfolded surface codes \cite{surface_surgery} and color codes \cite{color_surgery}.
Our goal is to highlight the balance between space (number of qubits) and time (circuit depth) resources and the distinction between physical and logical overhead.
We do not perform a detailed quantitative assessment and argue that it is premature without constraints provided by a mature qubit technology.

The reason to design new codes is to reduce the spacetime volume of quantum computations,
 but the relative value that we assign to space and time depends on details.
If space is the dominant constraint because of high manufacturing costs for qubits
 or spatial limits imposed by the limited size of vacuum chambers or dilution refrigerators,
 then the number of qubits is the design metric that we should minimize.
If time is the dominant constraint because we seek to reduce the length of quantum computations at any expense,
 then the circuit depth is the design metric that we should minimize.
Practically, the space and time requirements of logical qubit operations both depend on code distance $D$.
For two-dimensional topological codes of interest in this paper,
 a logical qubit requires an area containing $O(D^2)$ physical qubits
 and time for $O(D)$ rounds of error correction to operate a logical qubit reliably.
Codes have a physical overhead that sets the prefactors of these costs per logical operation,
 and a logical overhead corresponding to the number of elementary logical operations
 that are required to perform specific logical tasks.

\begin{figure}[!t]
\includegraphics{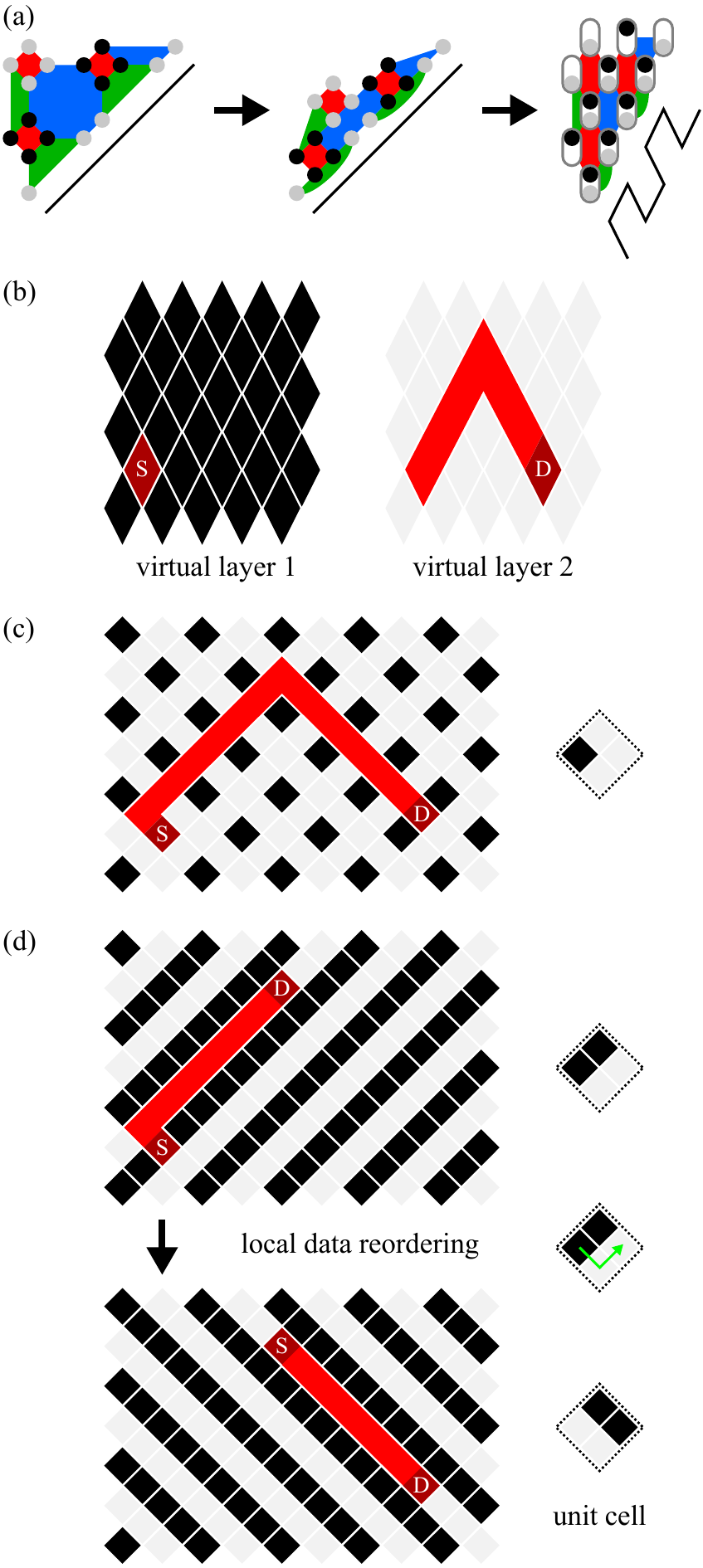}
\caption{\label{fig_layout} Examples of the space-time tradeoffs in logical qubit design.
(a) A distance-5 4.8.8 color code adapted to a virtual 2-layer qubit lattice with black (gray) qubits on the top (bottom) layer.
 It is more space efficient but less time efficient than the diamond surface code in Fig. \ref{fig_cone}.
Movement of a source logical qubit (S) to a destination (D) by code deformation on a lattice of diamond surface codes with
 (b) two virtual layers, which has the time efficiency of (c) a quarter-filled layer, and the space efficiency of (d) a half-filled layer.
}
\end{figure}

To better compare the surface and color codes, we adapt the most efficient 4.8.8 color code \cite{color_efficiency}
 to the virtual 2-layer lattice of qubits in Fig. \ref{fig_layout}a.
If we collapse the octagons to rectangles and snake the planar color code through both layers,
 then we can construct the stabilizer generators of the color code from products of stabilizer generators of two stacked surface codes.
In this configuration, the color code uses both layers, but only one quarter of the surface area of a diamond surface code with the same code distance.
Circuits similar to those in Sec. \ref{folded_QEC} can measure the stabilizer generators of a color code.
For the generators that combine two surface code generators,
 we can run two measurement circuits and omit the intermediate qubit measurement.
The error correction cycle of a color code thus requires twice as much time as a surface code.
By this simple analysis, a 4.8.8 color code and a diamond surface code of the same distance
 occupy the same spacetime volume and offer a factor of two tradeoff between space and time.

Color codes and folded surface codes both have transversal
 implementations of the single-qubit Clifford group that reduce
 their logical overhead relative to standard surface codes.
The transversal $X$, $Z$, $H$, and $S$ gates can be implemented between
 syndrome measurements to generate any of the 24 elements in the single-qubit Clifford group
 with a negligible cost relative to quantum error correction.
On a standard surface code, the $X$ and $Z$ gates are transversal,
 and the $H$ gate is transversal up to a geometric rotation of the logical qubit \cite{surface_surgery}.
These generate 8 elements of the Clifford group transversally up to rotations,
 but 12 elements require one $S$ gate and 4 elements require two $S$ gates.
The standard implementation of a logical $S$ gate on a surface code requires
 two logical ancilla qubits, one prepared in a stabilizer resource state \cite{surface_operations} and the other used for logical $CX$ gates \cite{surface_surgery}.
The total space-time volume of this logical $S$ gate is $\approx 72 D^3$ qubit rounds \cite{color_surgery},
 which overwhelms the cost of other single-qubit Clifford gates in resource estimates.

Folded surface codes also have an advantage over standard surface codes for the packing and movement of logical qubits on a grid of physics qubits.
With a free virtual layer of qubits, we can move logical qubits by extension and contraction but
 utilize only half of the data qubits as persistent logical qubits as shown in Fig. \ref{fig_layout}b.
The same operation on one layer utilizes only a quarter of the data qubits \cite{surface_surgery} in Fig. \ref{fig_layout}c.
However, a two-step movement operation in Fig. \ref{fig_layout}d can also utilize half of the data qubits.
This is another design choice with a factor of two tradeoff between space and time.

We are considering resource estimates as a function of code distance,
 but a more fair comparison is between codes of equal logical error rate.
The standard empirical scaling of the logical error rate $\epsilon_L$
 with physical error rate $\epsilon_P$ and code distance $D$ is 
\begin{equation}
 \epsilon_L \approx \epsilon_0 \left( \epsilon_P / \epsilon_T \right)^{\lfloor (D+1)/2 \rfloor}
\end{equation}
 for some offset $\epsilon_0$ and error threshold $\epsilon_T$.
With a depolarizing error channel applied to every gate, $\epsilon_T \approx 0.01$ for the surface code
 and $\epsilon_T \approx 0.001$ for the color code \cite{color_efficiency}.
This difference is caused by deeper syndrome measurement circuits and harder decoding problems for color codes.
If $\epsilon_P$ is between these $\epsilon_T$ values for some qubit technology, then only the surface code is viable.
If $\epsilon_P \ll \epsilon_T$, then the distance tradeoff of two codes with equivalent $\epsilon_L$ described by $\{ \epsilon_0 , \epsilon_T, D \}$ and $\{ \epsilon'_0 , \epsilon'_T, D' \}$ is
\begin{equation}
 \frac{D}{D'} \approx 1 + \frac{ \ln \epsilon_T - \ln \epsilon'_T }{\ln \epsilon_P} + O\left( \frac{1}{D} \right) + O\left( \left(\frac{\ln \epsilon_T}{\ln \epsilon_P}\right)^2 \right),
\end{equation}
 and differences in thresholds have a small effect.
Overall, the design space increases as the physical error rate decreases.

Beyond a simplistic physical error rate that assumes errors on qubits to be uniform and independent,
 there are numerous plausible scenarios that might distort resource estimation.
For example, if state preparation and measurement operations are significantly slower than unitary gates,
 then the differences in circuit depth between syndrome measurement circuits
 of the surface and color codes might have a weak effect on the total time and accumulated error of an error correction cycle.
The larger code capacity error thresholds become relevant if error accumulates mostly during measurement operations.
Another possible scenario is a prevalence of spatially correlated errors from control crosstalk.
Such errors might reduce the effective distance of a topological code to a ratio between the physical length of the code
 and an error correlation length.
The higher code efficiency of the color code relative to the surface code might then be negated by its spatial compactness.
We will not know what errors are practically relevant until a mature qubit technology has been subject to thorough characterization.

\section{Universal quantum computation\label{fusion}}

We now consider a topological implementation of the qubit fusion concept from Sec. \ref{fusion_subsection}
 with folded surface codes as a method for universal quantum computation.
The basic idea is to encode a four-dimensional logical Hilbert space as either one folded qudit surface code or a stack of two folded qubit surface codes.
With two codes, we are able to circumvent two important no-go theorems:
 that no single code can have a set of gates that are both transversal and universal \cite{transversal_nogo}
 and that no topological stabilizer code in two spatial dimensions can topologically protect a non-Clifford gate \cite{nonclifford_nogo}.
The two codes combined have transversal implementations of both qubit and qudit Clifford gates, which form a universal gate set.
The hard operation is converting between a qudit surface code and two qubit surface codes,
 which either requires the distillation of a logical resource state $|\overline{F}\rangle$
 or a new code conversion method.
 
\begin{figure}[!t]
\includegraphics{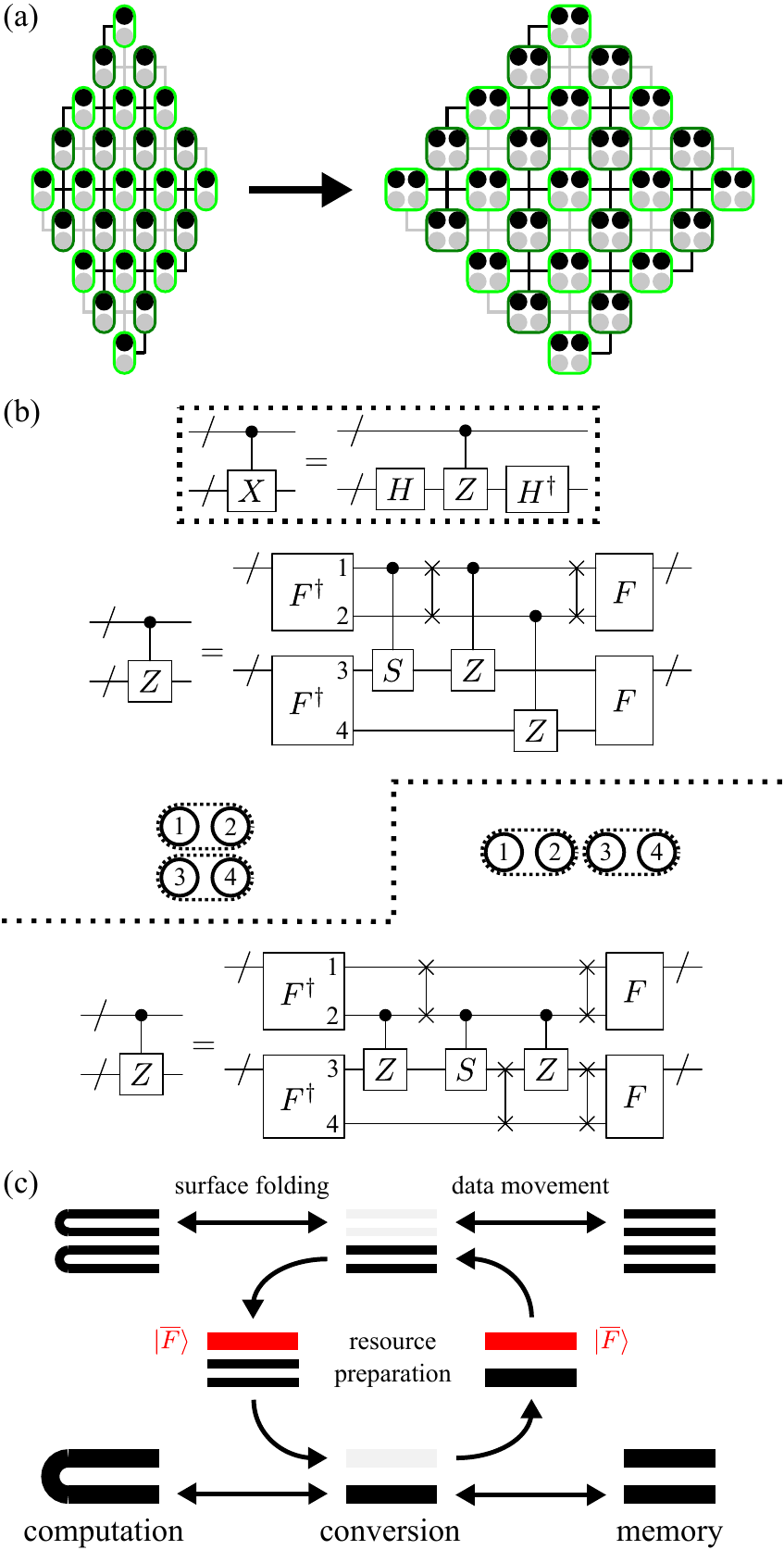}
\caption{\label{fig_uqc} Universal quantum computation on a planar qubit lattice.
(a) A stacked pair of four-dimensional qudit surface codes on a planar qudit lattice (left) can be implemented on a planar qubit lattice (right)
 by embedding each qudit in a pair of physical qubits.
 This layout of data and ancilla qubits in $2 \times 2$ clusters also can be operated as four stacked qubit surface codes.
 (b) 1-qudit Clifford gates in Fig. \ref{fig_fusion} can be implemented with nearest-neighbor 2-qubit gates, but the $CX$ gate is more complicated.
We decompose $CX$ gates into $H$ and $CZ$ gates (inset), and implement $CZ$ gates
 with nearest-neighbor 2-qubit gates that depend on the relative orientations of qubit pairs in every qudit.
These implementations further simplify if qubits are not returned to their initial orientation.
(c) Four virtual layers of surface codes have space to fold one qudit or two qubits (left), use a resource state for code conversion (center),
 or store two qudits or four qubits (right).
}
\end{figure}

We operate pairs of physical qubits as qudits to implement folded qudit surface codes on a qubit grid as shown in Fig. \ref{fig_uqc}.
The 2-qudit subsystems are now 4-qubit subsystems arranged in a square, which enables a virtual 4-layer qubit lattice.
We can use all of the qudit results from Sec. \ref{fold} by implementing qudit operations on neighboring qubit pairs in Fig. \ref{fig_uqc}b.
This qudit-in-qubit embedding has also been proposed recently for Hamiltonian-based topological quantum computation \cite{qudit_embedding}
 to implement a $\mathbb{Z}_4$ quantum double model \cite{quantum_double}.
In this language, we are implementing a $\mathbb{Z}_4 \times \mathbb{Z}_4$ quantum double model (the folded qudit surface code)
 and a $\mathbb{Z}_2 \times \mathbb{Z}_2 \times \mathbb{Z}_2 \times \mathbb{Z}_2$ quantum double model (the stacked pair of folded qubit surface codes)
 with nontrivial boundary conditions \cite{double_boundaries}.

Quantum error correction has been studied extensively for qubit surface codes
 and many of these results can be adapted to qudit surface codes.
The most important result is that the most likely error can be computed efficiently using minimum weight perfect matching
 with an error model of independent $X$ and $Z$ Pauli errors \cite{quantum_memory}.
This result also holds for the $d=4$ qudit surface code with an error model of independent $X$, $X^2$, $Z$, and $Z^2$ errors
 defined by applying four error channels,
\begin{align} \label{qudit_error}
 \mathcal{E}_X(\rho) &= (1-\epsilon_P) \rho + \epsilon_P X \rho X^\dag \notag \\
 \mathcal{E}_{X^2}(\rho) &= (1-\epsilon_P) \rho + \epsilon_P X^2 \rho X^2 \notag \\
 \mathcal{E}_Z(\rho) &= (1-\epsilon_P) \rho + \epsilon_P Z \rho Z^\dag \notag \\
 \mathcal{E}_{Z^2}(\rho) &= (1-\epsilon_P) \rho + \epsilon_P Z^2 \rho Z^2 .
\end{align}
This simple model factors the decoding problem by error type.
The parities of $X$ and $Z$ syndrome measurements are sensitive only to $Z$ and $X$ errors respectively,
 which enables them to be corrected separately from $Z^2$ and $X^2$ errors.
Error thresholds are identical to the qubit surface code in a code capacity model
 with errors applied between perfect rounds of error correction ($\epsilon_T \approx 0.1$)
 and in a phenomenological model that adds errors to syndrome measurements ($\epsilon_T \approx 0.03$) \cite{surface_threshold2}.
Both thresholds are bulk properties and are not altered by boundary conditions of folded surface codes.
Other error models have surface code thresholds that increase with qudit dimension \cite{decode_RG,qudit_threshold}.
This is consistent with our result because qubits and qudits have the same threshold but the error probability
 on a qudit ($\approx 4 \epsilon_P$) is higher than on a qubit ($\approx 2 \epsilon_P$) in our error model.
For a more realistic gate-based error model, thresholds will be lower for a folded qudit surface code
 than for an unfolded qubit surface code because of deeper syndrome measurement circuits.

The virtual 4-layer qubit lattice has several possible modes of operation.
In a ``memory" mode, we store a 16-dimensional logical Hilbert space as four qubits,
 two qudits, or two qubits and a qudit in a stack of surface codes.
We can apply strongly transversal $CX$ gates to any pair of these stacked codes.
In a ``computation" mode, we can apply transversal Clifford gates to stacked pairs of folded qubit surface codes or folded qudit surface codes.
This provides transversal access to the $11 \, 520$ elements of the 2-qubit Clifford group or the 768 elements of the 1-qudit Clifford group.
In a ``conversion" mode, we switch between a pair of qubit surface codes and a qudit surface code
 by consuming an $|\overline{F}\rangle$ on a stacked qudit surface code to apply an operation in Fig. \ref{fig_fusion}d. 
They can both be performed in place because partial qudit fission is transversal on surface codes.

With a steady supply of $|\overline{F}\rangle$ resources to consume, a stack of surface codes on a virtual 4-layer qubit lattice is a compact ``quantum logic unit."
Universal quantum computation within a 4-dimensional logical Hilbert space is performed by cycling between
 two logical qubits and one logical qudit to interleave qubit and qudit Clifford gates.
A unitary operation performed in this manner must be compiled into a Clifford+$F$ circuit.
A more conventional Clifford+$T$ quantum logic unit computing on two logical qubits can be implemented in the same amount of space
 with a supply of conventional magic state resources.
The relative value of Clifford+$T$ and Clifford+$F$ architectures depends on compiler efficiency and the cost of resource state preparation.
An $F$ gate can effectively apply more than one $T$ gate \cite{qubit_fusion}, 
 and a careful valuation will require the $T$-gate count and depth \cite{T_depth} for all elements of the 1-qudit Clifford group.
Only one protocol \cite{qubit_fusion} has been identified for $|\overline{F}\rangle$ distillation so far,
 and more research is needed to reach the same level of maturity as conventional magic state distillation.

A possible alternative to $|\overline{F}\rangle$ distillation is code conversion between two stacked qubit surface codes and a qudit surface code.
While such an operation has not yet been demonstrated,
 there are many similarities between the codes that hint at its feasibility.
First, we label qubit stabilizer generators by layer and separate qudit stabilizer generators into $X$, $X^2$, $Z$, and $Z^2$ types.
For an embedding consistent with Fig. \ref{fig_fusion},
 $X_2$-type and $Z_1$-type qubit generators are equivalent to $X^2$-type and $Z^2$-type qudit generators,
  and $Z_2$-type qubit generators commute with $Z$-type qudit generators.
Thus these codes share half of their generators, and three-quarters of their generators are mutually commuting.
As a result, the $X_2$-type and $Z_1$-type qubit logical strings are equivalent to the $X^2$-type and $Z^2$-type qudit logical strings.
With all of these similarities between codes, it is the $X_1$-type qubit generators and the $X$-type qudit generators that 
 are the primary source of difficulty in code conversion.

In addition to computational advantages, the virtual 4-layer qubit lattice benefits logical data movement over a lattice of stacked surface codes.
The lattice needs two empty layers to accommodate the simultaneous folding of all surface codes.
With one empty layer as in Fig. \ref{fig_layout}b, movement paths cannot overlap.
With two empty layers, we can overlap any pair of movement paths by assigning them locally to different layers.
This flexibility only applies to the movement of logical qubits,
 since logical qudits occupy two layers simultaneously.
Qudit movement is still useful to stack qudits for strongly transversal $CX$ gates
 that directly implement Toffoli gates as in Fig. \ref{fig_fusion}b.
This is useful for performing classical reversible logic, which is prevalent in the oracle functions of quantum algorithms.

\section{Conclusion\label{conclude}}

The main technical result of this paper is a construction of folded surface codes with transversal Clifford gates
 that is an alternative to a mapping from color codes \cite{fold_surface}.
We embed a virtual multilayer lattice of qubits on a physical planar lattice of qubits by clustering data and ancilla qubits.
This expands the design space of topologically protected logical qubits
 and enables new tradeoffs between space and time resources.

We combine folded surface codes on a virtual 4-layer qubit lattice
 with code conversion between a logical qubit pair and a logical qudit \cite{qubit_fusion}
 to form a framework for universal quantum computation in two spatial dimensions.
Abstractly, it protects two distinct forms of topological order with Abelian anyons rather than topological order with non-Abelian anyons,
 which is generally believed to be necessary for topological quantum computation in two spatial dimensions \cite{quantum_double}.
This framework still lacks efficient operations to switch orders either directly with code conversion or indirectly with resource distillation.
We are following a ``keystone" design principle whereby the most essential part is missing,
 and its future development is motivated by a compelling but incomplete design.
To compete with existing quantum computing proposals \cite{quantum_resources}, we will need
 basic principles for converting between qubit and qudit error correcting codes,
 more efficient protocols for $|\overline{F}\rangle$ distillation,
 efficient hybrid stabilizer simulations of qubits and qudits and their interconversion,
 efficient compilation of operations into Clifford+$F$ gates,
 and efficient scheduling of operations with geometric constraints of a virtual 4-layer lattice.
Ultimately, universal quantum computation is likely to coalesce around a universal gate set optimized for a planar lattice of qubits.

\begin{acknowledgments}
I thank Ciar\'{a}n Ryan-Anderson for useful discussions.
This work was supported by the Laboratory Directed Research and Development program at Sandia National Laboratories.
Sandia National Laboratories is a multi-program laboratory managed and  
operated by Sandia Corporation, a wholly owned subsidiary of Lockheed 
Martin Corporation, for the U.S. Department of Energy's National  
Nuclear Security Administration under contract DE-AC04-94AL85000.
\end{acknowledgments}

\end{document}